\documentclass[pra,twocolumn,showpacs]{revtex4}

\usepackage{graphics}       
\usepackage{bm}         
\usepackage{amsmath}        
\usepackage{latexsym}
\begin{document}
\title{Quantum Operations, State Transformations and Probabilities}
\author{Anthony Chefles}
\affiliation{Department of Physical Sciences, University of
Hertfordshire,
       Hatfield AL10 9AB, Hertfordshire, UK}

\begin{abstract}
\vspace{0.5cm} In quantum operations, probabilities characterise
both the degree of the success of a state transformation and, as
density operator eigenvalues, the degree of mixedness of the final
state.  We give a unified treatment of pure${\rightarrow}$pure
state transformations, covering both probabilistic and
deterministic cases.  We then discuss the role of majorization in
describing the dynamics of mixing in quantum operations.  The
conditions for mixing enhancement for all initial states are
derived.  We show that mixing is monotonically decreasing for
deterministic pure${\rightarrow}$pure transformations, and discuss
the relationship between these transformations and deterministic
LOCC entanglement transformations.
\\

\end{abstract}
\maketitle

\section{Introduction}
\renewcommand{\theequation}{1.\arabic{equation}}
\setcounter{equation}{0}

Information is carried by physical systems and encoded in their
states. The range of possible manipulations of information is thus
delimited by the scope of the set of possible operations on the
states of the signal carriers. It is for this reason that the
recent widespread fascination with the information-theoretic
properties of quantum systems\cite{NC} has been accompanied by a
renaissance in the study of the quantum operations formalism,
which determines what we can, and cannot do with the state of a
quantum system.

In quantum, as in classical information theory, the systems
considered may be in one of many possible states.  However,
quantum states can have attributes that have no exact classical
analogue, such as non-orthogonality and entanglement.  These
features of quantum states, together with the numerous, novel ways
in which quantum states can be manipulated, have given rise to
some intriguing discoveries in quantum information theory, such as
teleportation, classical capacity superadditivity and quantum
error correction.  Certain limitations on the way in which quantum
states can be manipulated, such as the no-cloning theorem, also
carry significant benefits, such as the security of quantum key
distribution and, relatedly, consistency with Special Relativity.

The many successes in determining optimal transformations for
carrying out specific important tasks, such as state
discrimination/estimation, approximate cloning and entanglement
manipulation, have led to some more general questions being asked
about the constraints imposed by the quantum formalism on state
manipulation. In this respect, Hardy and Song\cite{Hardysong} have
considered optimal universal manipulation of a qubit, while Alber,
Delgado and Jex\cite{Alber} have described universal bipartite
entanglement processes. Even more recently,
Fiur\'a\v{s}ek\cite{Fiurasek} has discussed the properties of
quantum operations which optimally approximate a given
transformation of one set of pure states into another with unit
probability.  The conditions under which such a transformation can
be carried out exactly, at least when the initial states are
linearly independent, have been derived in \cite{Medet}.

In this paper we continue to explore the properties of general
quantum operations and how they transform quantum states.
Section II is devoted to giving a unified treatment of
probabilistic and deterministic transformations between sets of
pure states.  We consider the following scenario: a quantum system
is prepared in one of the $N$ pure states
$|{\psi}^{1}_{j}{\rangle}$, where $j=1,{\ldots},N$. We wish to
implement the transformation
$|{\psi}^{1}_{j}{\rangle}{\rightarrow}|{\psi}^{2}_{j}{\rangle}$,
for some other set of $N$ pure states $|{\psi}^{2}_{j}{\rangle}$.
In general, the transformation will not be deterministic, and will
only succeed with some probability for each state.  We obtain
necessary and, for linearly independent initial states, sufficient
conditions for the existence of a quantum operation which carries
out this transformation for a fixed set of success probabilities.
We then examine some consequences of these conditions, and show
how they lead to simple derivations of established conditions for
deterministic state transformations and optimal unambiguous state
discrimination.

For a general quantum operation, when the initial state is pure,
the final state will often be mixed.  This effect is almost
ubiquitous and occurs under many circumstances where we wish to
preserve the information content of a quantum state, such as in
quantum communications and quantum computation. To understand this
mixing it helps to have an appreciation of its quantitative
features.  A suitable framework for the discussion of mixing is
provided by the concept of majorization.  This concept was
introduced by Uhlmann\cite{Uhlmann1,Uhlmann2,Uhlmann3} as a means
of quantifying mixing in density operators and probability
distributions and numerous useful theorems relating to
majorization have been discovered\cite{Bhatia}.

The subject of majorization has  recently received renewed
attention in quantum information theory, mainly as result
Nielsen's discovery that it provides a suitable framework for the
discussion pure, bipartite entanglement
transformations\cite{Nielsen1}.  More recently,
Nielsen\cite{Nielsen2} has derived several interesting
majorization relations for static and dynamic mixing of quantum
states, latterly in association with generalised measurements (see
also the related analysis by Fuchs and Jacobs\cite{Fuchs}.)
Nielsen has also showed that a density operator can represent some
ensemble of pure states with fixed probabilities if and only if a
certain majorization relation is satisfied\cite{Nielsen3}.

In section III, we describe and employ the concept of majorization
as a tool to help us understand the dynamics of mixing in quantum
operations. All nonunitary quantum operations transform (at least
some) pure states into mixed states.  This begs the question:
under what conditions does a quantum operation never decrease the
extent to which any initial state is mixed? Majorization is a
suitable tool for comparing the degree of disorder in the initial
and final states, and a sufficient condition for this monotonic
mixing was derived, in purely algebraic context, by Bapat and
Sunder\cite{Bapat}.  We give a simple derivation of their
condition within the context of quantum operations, and show that
this condition is also in fact a necessary condition.  We then
examine majorization in relation to deterministic pure state
transformations, and derive an intuitive and
information-theoretically satisfying majorization relation for
such operations.

\section{Transformations between sets of pure states}
\renewcommand{\theequation}{2.\arabic{equation}}
\setcounter{equation}{0}
\subsection{Transformation conditions for fixed probabilities}
Consider the following situation: we have in our possession a
quantum system with a finite, $D$ dimensional Hilbert space ${\cal
H}$.  The initial state of the system is pure, and is an element
of the set $\{|{\psi}_{j}^{1}{\rangle}\}$, where $j=1,{\ldots},N$
for some finite $N$.  Our aim is to implement a probabilistic
transformation ${\cal P}$ which transforms the state
$|{\psi}_{j}^{1}{\rangle}$ into some other pure state
$|{\psi}_{j}^{2}{\rangle}$ for each $j$.

It is well known, from studies of particular transformations such
as unambiguous state discrimination\cite{Linear} and probabilistic
cloning\cite{Duanguo} , that we cannot in general expect the
probability of success to be equal to 1. Let $p_{j}$ then be the
probability of successful transformation of
$|{\psi}_{j}^{1}{\rangle}$ into $|{\psi}_{j}^{2}{\rangle}$. These
probabilities may be represented as the components of a vector
${\mathbf p}=\{p_{j}\}$.

Generally speaking, the transformation ${\cal P}$ will be
represented by a completely-positive, linear map. We would like to
be able to determine unambiguously whether or not the desired
transformation has succeeded.  This requirement implies that the
procedure will have two possible outcomes: success or failure. It
will be described by the transformation operators $\{A_{kr}\}$,
where $r=S,F$, corresponding to success and failure respectively,
and $k=1,{\ldots},M$, for some $M$. If the system is prepared in a
state represented by an initial density operator ${\rho}$, then
the probability of the $r$th outcome is determined by the positive
quantum detection operator, or positive operator-valued measure
(POVM) element
\begin{equation}
E_{r}=\sum_{k}A^{\dagger}_{kr}A_{kr }.
\end{equation}
Throughout this paper, when we speak of a positive operator or
matrix, we will, unless otherwise indicated, mean one which is
positive semidefinite. The probability of the $r$th outcome is
given by
\begin{equation}
p_{r}({\rho})={\mathrm{Tr}}{\rho}E_{r},
\end{equation}
where
\begin{equation}
\sum_{r}E_{r}=1
\end{equation}
The post-measurement state corresponding to the $r$th outcome is
\begin{equation}
{\rho}_{r}=\sum_{k}A_{kr}{\rho}A^{\dagger}_{kr }/p_{r}({\rho}).
\end{equation}
It is clear from Eq. (2.1) that $E_{r}$ is positive.  From the
resolution of the identity in Eq. (2.3) we see that
\begin{equation}
0{\leq}E_{r}{\leq}1.
\end{equation}
Let us denote by ${\Sigma}_{\mathbf p}({\cal P})$ the set of
admissible probability vectors for this transformation ${\cal P}$.
We would like to determine the conditions under which a particular
probability vector is an element of ${\Sigma}_{\mathbf p}({\cal
P})$. The necessary and sufficient conditions for the existence of
a transformation which succeeds with probability vector ${\mathbf
p}{\in}{\Sigma}_{\mathbf p}({\cal P})$ are that it can be realised
by a set of linear transformation operators as in Eq. (2.4)
 and that the
corresponding POVM element satisfies Eq. (2.1).  These criteria,
while correct, may not always be the most helpful, due to the
large number of parameters describing the transformation
operators.  The following theorem gives simpler necessary and, for
linearly independent initial states, sufficient conditions for the
existence of such a transformation.

\newtheorem{theorem}{Theorem}
\begin{theorem} Let $\{|{\psi}_{j}^{1}{\rangle}\}$ be a set of $N$ pure quantum states
spanning a D dimensional Hilbert space ${\cal H}$.  Let
$\{|{\psi}_{j}^{2}{\rangle}\}$ be another set of $N$ pure states
lying in ${\cal H}$.  Let the Gram matrices of the initial and
final sets be denoted by ${\mbox{\boldmath ${\Gamma}$}}_{1}$ and
${\mbox{\boldmath ${\Gamma}$}}_{2}$ respectively.  If there exists
a probabilistic transformation ${\cal
P}:\{|{\psi}_{j}^{1}{\rangle\}}{\rightarrow}\{|{\psi}_{j}^{2}{\rangle}\}$
with probability vector ${\mathbf p}$, then there exists an
$N{\times}N$ matrix ${\mbox{\boldmath ${\Pi}$}}$ which satisfies
the following conditions:
\begin{eqnarray}
 \nonumber &(1.{\mathrm{a}})&\;{\mbox{\boldmath
${\Pi}$}}{\geq}0, \\
 \nonumber &(1.{\mathrm{b}})&\;{\mathrm{Diag}}({\mbox{\boldmath
${\Pi}$}})={\mathbf p},
\\
 \nonumber &(1.{\mathrm{c}})&\;{\mbox{\boldmath ${\Gamma}$}}_{1}-{\mbox{\boldmath
${\Pi}$}}\circ{\mbox{\boldmath ${\Gamma}$}}_{2}{\geq}0,
\end{eqnarray}
where `${\circ}$' denotes the Hadamard (or Schur) matrix product.
These conditions are also sufficient if the set
$\{|{\psi}_{j}^{1}{\rangle}\}$ is linearly independent.
\end{theorem}

Prior to giving a proof of this theorem, we recall that the
$N{\times}N$ Gram matrix ${\mbox{\boldmath
${\Gamma}$}}=\{{\gamma}_{j'j}\}$ corresponding to a set of $N$
pure states $|{\psi}_{j}{\rangle}$ has elements
${\gamma}_{j'j}={\langle}{\psi}_{j'}|{\psi}_{j}{\rangle}$.  Also,
the Hadamard product ${\mathbf A}{\circ}{\mathbf B}$ of two
matrices  ${\mathbf A}=\{a_{j'j}\}$ and ${\mathbf B}=\{b_{j'j}\}$
has $j'j$ element $a_{j'j}b_{j'j}$.
\\

{\noindent}{\bf Proof:}  We begin by proving the necessity part of
this theorem.  To do this, we note that there must exist complex
coefficients $c_{kj}$ such that
\begin{equation}
A_{kS}|{\psi}^{1}_{j}{\rangle}=c_{kj}|{\psi}^{2}_{j}{\rangle}.
\end{equation}
We can consider these coefficients to be the elements of an
$M{\times}N$ matrix ${\mathbf C}=\{c_{kj}\}$.  Let us now
introduce the $N{\times}N$ matrix ${\mbox{\boldmath
${\Pi}$}}=\{{\pi}_{j'j}\}$ defined by
\begin{equation}
{\mbox{\boldmath ${\Pi}$}}={\mathbf C}^{\dagger}{\mathbf C}.
\end{equation}
This matrix is clearly positive and thus satisfies (1.a). To see
that it also satisfies condition (1.b), we make use of the fact
that
$p_{j}={\langle}{\psi}^{1}_{j}|\sum_{k}A^{\dagger}_{kS}A_{kS}|{\psi}^{1}_{j}{\rangle}=\sum_{k}|c_{kj}|^{2}$.
This is easily shown to be equal to ${\pi}_{jj}$ using Eq. (2.7),
which implies that ${\mbox{\boldmath ${\Pi}$}}$ satisfies
condition (1.b).  Finally, condition (1.c) can be verified by
imposing (2.5), which requires the expectation value of
$\sum_{k}A^{\dagger}_{kS}A_{kS}$ to be no greater than 1 for any
state. Consider an arbitrary pure state $|{\phi}{\rangle}$ in the
subspace spanned by the $\{|{\psi}_{j}^{1}{\rangle}\}$.  We may
write it is
$|{\phi}{\rangle}=\sum_{j}v_{j}|{\psi}_{j}^{1}{\rangle}$, and
calculate
\begin{eqnarray}
{\langle}{\phi}|\left[\sum_{k}A^{\dagger}_{kS}A_{kS}\right]|{\phi}{\rangle}&=&\sum_{jj'}v^{*}_{j'}v_{j}{\pi}_{j'j}{\gamma}_{j'j}^{2},
\nonumber \\
&{\leq}&{\langle}{\phi}|{\phi}{\rangle}=\sum_{jj'}v^{*}_{j'}v_{j}{\gamma}_{j'j}^{1}.
\end{eqnarray}
The requirement that $\sum_{k}A^{\dagger}_{kS}A_{kS}{\leq}1$ is
then seen to be equivalent to the inequality
\begin{equation}
\sum_{jj'}v^{*}_{j'}v_{j}({\pi}_{j'j}{\gamma}_{j'j}^{2}-{\gamma}^{1}_{j'j}){\leq}0,
\end{equation}
which holds for every vector ${\mathbf v}$.  From this it follows
that the $N{\times}N$ matrix with elements
$\{{\gamma}^{1}_{j'j}-{\pi}_{j'j}{\gamma}_{j'j}^{2}\}$ is
positive, which is exactly what is expressed, more concisely, by
condition (1.c).

To prove the converse for linearly independent initial states, we
assume the existence of a matrix ${\mbox{\boldmath
${\Pi}$}}=\{{\pi}_{j'j}\}$ which satisfies the three conditions in
(1.a)-(1.c). Positivity enables us to factorise ${\mbox{\boldmath
${\Pi}$}}$ as ${\mathbf C}^{\dagger}{\mathbf C}$, for some
$M{\times}N$ matrix ${\mathbf C}=\{c_{kj}\}$, where the integer
$M$ may take any value not less than $N$.  Let us now define the
transformation operators

\begin{equation}
A_{kS}=\sum_{j}\frac{c_{kj}}{{\langle}{\tilde
\psi}^{1}_{j}|{\psi}^{1}_{j}{\rangle}}|{\psi}^{2}_{j}{\rangle}{\langle}{\tilde
\psi}^{1}_{j}|.
\end{equation}
The $|{\tilde \psi}^{1}_{j}{\rangle}$ are the reciprocal vectors
corresponding to the states $|{\psi}^{1}_{j}{\rangle}$. These have
been found, in studies of operations of unambiguous state
discrimination\cite{Linear} and deterministic
transformations\cite{Medet}, to be extremely useful in dealing
with transformations of sets of linearly independent states.  The
state $|{\tilde \psi}^{1}_{j}{\rangle}$ is defined as that in
${\cal H}$ which is orthogonal to all $|{\psi}_{j'}^{1}{\rangle}$
for $j{\neq}j'$ and is, up to a phase, unique.

From the definition, we see that (1.a) is automatically satisfied.
Also making use of Eq. (2.2), is clear that $p_{j}$, the
transformation probability for the $j$th state, given by
${\langle}{\psi}_{j}|\sum_{k}A^{\dagger}_{kS}A_{kS}|{\psi}_{j}{\rangle}={\pi}_{jj}$.
This shows that condition (1.b) is satisfied.  Finally, the
necessary and sufficient condition for the transformation
operators in Eq. (2.10) to be physically realisable is that
$\sum_{k}A^{\dagger}_{kS}A_{kS}{\leq}1$.  If condition (1.c) is
satisfied, then so is inequality (2.5), which is equivalent to
$\sum_{k}A^{\dagger}_{kS}A_{kS}{\leq}1$.  This completes the
proof.${\Box}$\\

\subsection{Examples}
It is instructive to see how established results relating to
specific transformations follow from the general conditions in
Theorem 1.  The first kind of transformation we shall consider is
a deterministic transformation, where all of the $p_{j}$ are equal
to 1.  Let us write ${\mathbf G}={\mbox{\boldmath
${\Gamma}$}}_{1}-{\mbox{\boldmath ${\Pi}$}}{\circ}{\mbox{\boldmath
${\Gamma}$}}_{2}$.  As a consequence of (1.c),  ${\mathbf G}$ must
be positive.  The diagonal elements of ${\mbox{\boldmath
${\Gamma}$}}_{1}$,${\mbox{\boldmath ${\Gamma}$}}_{2}$ and, as a
consequence of (1.c), ${\mbox{\boldmath ${\Pi}$}}$ are all equal
to 1.  It follows that the diagonal elements, and hence the trace,
of ${\mathbf G}$ are equal to zero. The only positive matrix with
zero trace is the zero matrix. Therefore,

\begin{equation}
{\mbox{\boldmath ${\Gamma}$}}_{1}-{\mbox{\boldmath
${\Pi}$}}{\circ}{\mbox{\boldmath ${\Gamma}$}}_{2}=0.
\end{equation}
One situation which is of particular interest is that which arises
when  ${\mbox{\boldmath ${\Gamma}$}}_{2}$ has no zero elements,
which corresponds to all of the final states being non-orthogonal.
When this is so, we can conclude that
\begin{equation}
{\mbox{\boldmath ${\Pi}$}}={\mbox{\boldmath
${\Gamma}$}}_{1}{\circ}{\mbox{\boldmath
${\Gamma}$}}_{2}^{{\circ}-1},
\end{equation}
where ${\mbox{\boldmath ${\Gamma}$}}_{2}^{{\circ}-1}$ is the
Hadamard inverse of ${\mbox{\boldmath ${\Gamma}$}}_{2}$.  The
Hadamard inverse of a matrix ${\mathbf A}=\{a_{j'j}\}$ has
elements $1/a_{j'j}$.  Finally, imposing condition (1.b) gives
\begin{equation}
{\mbox{\boldmath ${\Gamma}$}}_{1}{\circ}{\mbox{\boldmath
${\Gamma}$}}_{2}^{{\circ}-1}{\geq}0,
\end{equation}
which is identical to condition (ii) in \cite{Medet} for a
deterministic transformation expressed in terms of Gram matrices
and Hadamard product notation.

The second case we shall consider is that of unambiguous state
discrimination.  Here, the final set of states is an orthonormal
set, and so ${\mbox{\boldmath ${\Gamma}$}}_{2}={\mathbf 1}$.  Let
${\mbox{\boldmath ${\Delta}$}}({\mathbf p})$ be the matrix with
$j'j$ element $p_{j}{\delta}_{j'j}$.  Then ${\mbox{\boldmath
${\Pi}$}}\circ{\mbox{\boldmath ${\Gamma}$}}_{2}={\mbox{\boldmath
${\Delta}$}}({\mathbf p})$.  Inserting this into (1.c) gives the
inequality
\begin{equation}
{\mbox{\boldmath ${\Gamma}$}}_{1}-{\mbox{\boldmath
${\Delta}$}}({\mathbf p}){\geq}0.
\end{equation}
This is precisely the inequality obtained by Duan and Guo using a
unitary-reduction scheme\cite{Duanguo}.

For a probability vector ${\mathbf p}$ which satisfies this
inequality, the corresponding ${\mbox{\boldmath ${\Pi}$}}$ may be
assumed to take a particularly simple form.  If
${\pi}_{j'j}=\sqrt{p_{j'}p_{j}}$, then it can easily be shown that
${\mbox{\boldmath ${\Pi}$}}$ satisfies both conditions (1.a) and
(1.b), and that (1.c) is equivalent to (2.14).  This
${\mbox{\boldmath ${\Pi}$}}$ matrix is clearly proportional to a
rank-one projector.

Matrices of this form have an interesting significance in relation
to the following question: under what additional conditions can
the transformation ${\cal P}$ be carried out with probability
vector ${\mathbf p}$ when only one of the $A_{kS}$ is non-zero?
That is, we are interested in implementing the transformation
using with just two transformation operators, $A_{S}$ and $A_{F}$,
respectively implementing and failing to implement the
transformation, and satisfying
$A_{S}^{\dagger}A_{S}+A_{F}^{\dagger}A_{F}=1$. Here we shall show
that the necessary and, for linearly independent initial states,
sufficient condition for the transformation ${\cal P}$ be to be
implementable this way with probability vector ${\mathbf p}$ is
that there exists a matrix ${\mbox{\boldmath ${\Pi}$}}$ which, in
addition to satisfying conditions (1.a)-(1.c) above, is also
proportional to a rank-one projector.

To prove necessity, we observe that for the transformation to meet
our specifications, there must exist some operator $A_{S}$ such
that
\begin{equation}
A_{S}|{\psi}^{1}_{j}{\rangle}=c_{j}|{\psi}^{2}_{j}{\rangle},
\end{equation}
for some coefficients $c_{j}$.  Let us now define the $N{\times}N$
matrix ${\mbox{\boldmath ${\Pi}$}}=\{{\pi}_{j'j}\}$ where
${\pi}_{j'j}=c^{*}_{j'}c_{j}$. This matrix is clearly proportional
to a rank-one projector.  The proof that this matrix must satisfy
the three conditions of Theorem 1 proceeds as in the more general
case.  It is clearly positive, and so satisfies condition (1.a).
Condition (1.b) follows from the fact that Eq. (2.15) gives
\begin{equation}
p_{j}=|c_{j}|^{2},
\end{equation}
and the derivation of condition (1.c) is essentially identical to
that of the more general case; obtaining it amounts to nothing
more than dropping the index $k$.  This proves necessity.

To prove sufficiency for linearly independent states, let
${\mbox{\boldmath ${\Pi}$}}$ be an $N{\times}N$ matrix
proportional to a rank-one projector. It follows that
${\pi}_{j'j}=c^{*}_{j'}c_{j}$ for some $c_{j}$. With these
coefficients, we construct the operator
\begin{equation}
A_{S}=\sum_{j}\frac{c_{j}}{{\langle}{\tilde
\psi}^{1}_{j}|{\psi}^{1}_{j}{\rangle}}|{\psi}^{2}_{j}{\rangle}{\langle}{\tilde
\psi}^{1}_{j}|.
\end{equation}

The remainder of the proof proceeds as in the more general case.
Clearly
$A_{S}|{\psi}_{j}^{1}{\rangle}=c_{j}|{\psi}_{j}^{2}{\rangle}$, as
is required.  The success probability for the $j$th state is
$p_{j}={\langle}{\psi}_{j}^{1}|A_{S}^{\dagger}A_{S}|{\psi}_{j}^{1}{\rangle}={\pi}_{jj}=|c_{j}|^{2}$.
We can finally make use of condition (1.c) as before to show that
$E_{S}=A_{S}^{\dagger}A_{S}{\leq}1$.

\section{Quantum operations and majorization}
\renewcommand{\theequation}{3.\arabic{equation}}
\subsection{Majorization relations and mixing}
So far we have been considering quantum operations which convert
one set of pure states into another, either deterministically or
probabilistically.  It is well-known, however, that a typical
operation will convert pure states into mixed states.  This effect
is often undesirable.  For example, one of the principle current
obstacles in the way of realising quantum computers is the
phenomenon of decoherence, which is the mixing of the state of the
computer by unwanted, uncontrollable environmental influences.

The mixing of quantum states is intimately connected with
entanglement.  In this example, decoherence arises due to the
entanglement of the computer with the environment.  If two systems
become entangled, their individual states will be mixed even
though the state of the entire system may remain pure.

It follows from this that measures of entanglement and mixedness
ought to be intimately related, at least when the entire system is
a pure, bipartite state.   Indeed, the von Neumann entropy of one
of subsystems simultaneously satisfies many of the natural
requirements of an entanglement measure and also those of a
measure of how mixed a subsystem state is.  However, being a
single quantity, it is unable to quantify many specific details of
entanglement or mixedness, in much the same way that the Shannon
entropy of a source in classical information theory, while being
sufficient to describe many important things, like the maximum
asymptotically error-free transmission rate, is a less complete
description of the source than the source symbols accompanied with
their respective {\em a priori} probabilities.

In the study of pure, bipartite entanglement, the analogous, more
complete description is given by the eigenvalues of the subsystem
density operators.  The prominence of these quantities becomes
apparent when the relationship between entanglement and
deterministic local operations with classical communication (LOCC)
is taken into consideration. Entanglement is nonincreasing under
such operations.  This implies that, if one state
$|{\Psi}_{1}{\rangle}$ can be transformed into another state
$|{\Psi}_{2}{\rangle}$  by deterministic LOCC, then
$|{\Phi}_{2}{\rangle}$ can be no more entangled than
$|{\Psi}_{1}{\rangle}$ with respect to any reasonable entanglement
measure.  The role of the subsystem density operator eigenvalues
in determining the conditions under which such a transformation is
possible was made clear by Nielsen\cite{Nielsen1}, who showed that
the necessary and sufficient condition for such a transformation
to be possible is a simple majorization relation.

In view of this and the connection between entanglement and mixing
of subsystem states, we should expect majorization to play a
similarly important role in describing mixedness.  Indeed, that
this is so was understood by
Uhlmann\cite{Uhlmann1,Uhlmann2,Uhlmann3} who originated the
concept, motivated by the problem of finding a universal framework
for the quantification of mixing.

It would be helpful, given current concerns about issues such as
decoherence, to understand the mixing properties of quantum
operations.  Majorization provides an eminently suitable framework
for the discussion of this issue, and our aim is to use it to help
us understand the information loss, which often occurs in quantum
operations and manifests itself as mixing.  Fortunately, some
progress has been made in this direction.  Some intriguing
theorems in linear algebra due to Bapat and Sunder\cite{Bapat} are
particularly useful in this context. Here, we will employ, and
indeed slightly enhance one of their results within the framework
of quantum operations, to obtain the necessary and sufficient
condition for a quantum operation to increase mixing, in terms of
majorization, for every initial state.

We then give an intuitive information-theoretic argument that the
density operator for a pure state ensemble should not become more
mixed when the pure states undergo a deterministic transformation
into another set of pure states, and that the majorization
relation we have hitherto considered ought not to apply (except in
a certain extremal, indeed trivial case) under such circumstances.
We then prove that, in fact, it is precisely the reverse
majorization relation that is always true.

Prior to doing so, we will briefly review the relevant concepts.
Consider two $N$ component vectors ${\mbox{\boldmath
${\lambda}$}}=\{{\lambda}_{r}\}$ and ${\mbox{\boldmath
${\sigma}$}}=\{{\sigma}_{r}\}$. The components will be taken to be
real and positive.  From these vectors, we construct two further
vectors ${\mbox{\boldmath
${\lambda}$}}^{\downarrow}=\{{\lambda}^{\downarrow}_{r}\}$ and
${\mbox{\boldmath
${\sigma}$}}^{\downarrow}=\{{\sigma}^{\downarrow}_{r}\}$.  The
components of ${\mbox{\boldmath ${\lambda}$}}^{\downarrow}$ and
${\mbox{\boldmath ${\sigma}$}}^{\downarrow}$ are those of
${\mbox{\boldmath ${\lambda}$}}$ and ${\mbox{\boldmath
${\sigma}$}}$ arranged in decreasing order.  The vector
${\mbox{\boldmath ${\lambda}$}}$ is said to {\em majorize} the
vector ${\mbox{\boldmath ${\sigma}$}}$ iff the following
conditions hold:
\begin{eqnarray}
\sum_{r=1}^{k}{\sigma}^{\downarrow}_{r}&{\leq}&\sum_{r=1}^{k}{\lambda}^{\downarrow}_{r},
\;\;\;\;\;\;1{\leq}k{\leq}N-1,
\\
\sum_{r=1}^{N}{\sigma}^{\downarrow}_{r}&=&\sum_{r=1}^{N}{\lambda}^{\downarrow}_{r}.
\end{eqnarray}
This majorization of ${\mbox{\boldmath ${\sigma}$}}$ by
${\mbox{\boldmath ${\lambda}$}}$ is written as ${\mbox{\boldmath
${\sigma}$}}{\prec}{\mbox{\boldmath ${\lambda}$}}$.

In the context of probability of the vectors ${\mbox{\boldmath
${\sigma}$}}$ and ${\mbox{\boldmath ${\lambda}$}}$ are probability
distributions, satisfying
$\sum_{r}{\sigma}_{r}=\sum_{r}{\lambda}_{r}=1$. The majorization
relation ${\mbox{\boldmath ${\sigma}$}}{\prec}{\mbox{\boldmath
${\lambda}$}}$ says that the distribution ${\mbox{\boldmath
${\sigma}$}}$ is no less mixed than ${\mbox{\boldmath
${\lambda}$}}$.  Two identities relating to majorization will be
of particular importance in what follows.  These are \\

\noindent (i) The vectors ${\mbox{\boldmath ${\sigma}$}}$ and
${\mbox{\boldmath ${\lambda}$}}$ satisfy the majorization relation
${\mbox{\boldmath ${\sigma}$}}{\prec}{\mbox{\boldmath
${\lambda}$}}$ if and only if there is a doubly stochastic matrix
${\mathbf S}$ such that ${\mbox{\boldmath ${\sigma}$}}={\mathbf
S}{\mbox{\boldmath ${\lambda}$}}$.  A doubly stochastic matrix is
a matrix whose elements are real, non-negative, and where the sum
of the elements in each row and column is equal to 1. \\

\noindent (ii) If ${\mbox{\boldmath
${\sigma}$}}{\prec}{\mbox{\boldmath ${\lambda}$}}$, and
${\lambda}_{r}=1/N$, then  ${\sigma}_{r}=1/N$ also.  This is
effectively a statement of the fact that if a probability
distribution ${\mbox{\boldmath ${\sigma}$}}$ is no less mixed than
another probability distribution ${\mbox{\boldmath ${\lambda}$}}$,
and ${\mbox{\boldmath ${\lambda}$}}$ is the maximally mixed, or
uniform distribution, then ${\mbox{\boldmath ${\sigma}$}}$ must
also be the uniform distribution.
\\
\subsection{Mixing enhancement and trace-preserving maps}
Here, we will employ majorization as a tool to help us understand
the increase of disorder in the state of a system which occurs in
many quantum operations.  We will consider a quantum system
prepared initially in the state ${\rho}_{1}$ which then undergoes
the transformation

\begin{equation}
{\rho}_{1}{\rightarrow}{\rho}_{2}=\sum_{k=1}^{M}A_{k}{\rho}_{1}A_{k}^{\dagger},
\end{equation}
where
\begin{equation}
\sum_{k=1}^{M}A^{\dagger}_{k}A_{k}=1.
\end{equation}
The degree of mixedness of a quantum state is completely
characterised by the density operator eigenvalues.  The vector of
eigenvalues of a density operator ${\rho}$ will be denoted by
${\mbox{\boldmath ${\lambda}$}}({\rho})$.  When is it true that
${\mbox{\boldmath ${\lambda}$}}({\rho}_{2}){\prec}{\mbox{\boldmath
${\lambda}$}}({\rho}_{1})$, the final state ${\rho}_{2}$ can be
characterised as being at least as mixed as the initial state
${\rho}_{1}$.

It is not true that for every quantum operation, the final state
will always be at least as mixed as the initial state, for every
initial state. For example, suppose that we carry out a von
Neumann measurement in the orthonormal basis
$\{|x_{k}{\rangle}\}$, and when we obtain result $k$, carry out a
unitary transformation which converts the state $|x_{k}{\rangle}$
into some pure state $|x{\rangle}$. For this procedure, the final
state will be the pure state $|x{\rangle}$, irrespective the
initial state, and how mixed it is.  An operation of this kind,
which may be viewed as an idealised kind of state preparation
procedure, clearly does not increase mixedness.

The following question then arises: under what conditions does a
trace-preserving quantum operation always increase mixedness or
disorder in the sense of majorization, for every initial state?
The answer is given by

\begin{theorem} Consider a completely positive, linear, trace-preserving map described by Eqs. (3.20) and (3.21).  The eigenvalues of the initial density operator majorize those of
the final density operator, that is
\begin{equation}
{\mbox{\boldmath ${\lambda}$}}({\rho}_{2}){\prec}{\mbox{\boldmath
${\lambda}$}}({\rho}_{1}),
\end{equation}
for every initial density operator ${\rho}_{1}$ if, and only if,
\begin{equation}
\sum_{k}A_{k}A^{\dagger}_{k}=1.
\end{equation}
\end{theorem}
{\noindent}{\bf Proof:} The sufficiency part of this theorem comes
from a more general result due to Bapat and Sunder\cite{Bapat},
and we will establish it by a variation on the relevant parts of
their argument.  Numerous extensions and consequences of their
work are discussed by Visick\cite{Visick}. Let
$\{|{\phi}_{r}^{1}{\rangle}\}$ and $\{|{\phi}_{r'}^{2}{\rangle}\}$
be complete, orthonormal sets of eigenvectors of ${\rho}_{1}$ and
${\rho}_{2}$ respectively. If either density operator has zero
eigenvalues, then we simply complete the orthonormal basis with an
orthonormal set spanning the kernel. From Eq. (3.20), we obtain
\begin{equation}
{\mbox{\boldmath ${\lambda}$}}({\rho}_{2})={\mathbf
S}{\mbox{\boldmath ${\lambda}$}}({\rho}_{1}),
\end{equation}
where we have defined the matrix ${\mathbf S}=\{S_{r'r}\}$ with
elements
\begin{equation}
S_{r'r}=\sum_{k}|{\langle}{\phi}_{r'}^{2}|A_{k}|{\phi}_{r}^{1}{\rangle}|^{2}.
\end{equation}
Clearly, $S_{r'r}$ is real and nonnegative.  The majorization
relation (3.22) will hold for every initial density operator
${\rho}_{1}$ if ${\mathbf S}$ is doubly stochastic, which will be
the case if the row and column sums of ${\mathbf S}$ are equal to
one. For the row sum, we have
\begin{equation}
\sum_{r'}S_{r'r}={\langle}{\phi}_{r}^{1}|\left[\sum_{k}A^{\dagger}_{k}A_{k}\right]|{\phi}_{r}^{1}{\rangle}=1,
\end{equation}
as a consequence of the completeness of the
$\{|{\phi}_{r'}^{2}{\rangle}\}$ and the resolution of the identity
in Eq. (3.21).  For the column sum, we see that
\begin{equation}
\sum_{r}S_{r'r}={\langle}{
\phi}_{r'}^{2}|\left[\sum_{k}A_{k}A^{\dagger}_{k}\right]|{
\phi}_{r'}^{2}{\rangle}=1,
\end{equation}
when Eq. (3.23) holds, where we have used the completeness of the
$\{|{\phi}_{r}^{1}{\rangle}\}$.  So, when Eq. (3.23) is true, the
matrix ${\mathbf S}$ is doubly stochastic and the majorization
relation in (3.22) holds for every initial density operator
${\rho}_{1}$. This proves sufficiency.

To prove necessity, we must show that Eq. (3.23) follows if the
majorization relation (3.22) is true for every initial density
operator ${\rho}_{1}$.   Actually, we need only consider the case
when ${\rho}_{1}$ is the maximally mixed state, that is,
${\rho}_{1}=1/D$, which implies that
${\lambda}_{r}({\rho}_{1})=1/D$. This, together with the identity
(ii), suffices to determine the final density operator
${\rho}_{2}$ completely.  As a consequence of identity (ii), the
only possible choice for ${\mbox{\boldmath
${\lambda}$}}({\rho}_{2})$ which is consistent with
${\lambda}_{r}({\rho}_{1})=1/D$ is ${\mbox{\boldmath
${\lambda}$}}({\rho}_{2})={\mbox{\boldmath
${\lambda}$}}({\rho}_{1})$.  It follows that ${\rho}_{2}$ must
also be the maximally mixed state. Inserting
${\rho}_{1}={\rho}_{2}=1/D$ into Eq. (3.20), and multiplying both
sides by $D$ immediately gives Eq. (3.23), completing the
proof.${\Box}$\\

Condition (3.23) is always satisfied if the $A_{k}$ are normal
operators, which is a sufficient condition for the sums in (3.21)
and (3.23) to be identical. It follows that any generalised
measurement described by a POVM with elements $E_{k}$ will satisfy
(3.23) if we choose the transformation operators to be
$A_{k}=\sqrt{E_{k}}$,. This choice of transformation operators for
a generalised measurement has been termed the `rawest'
implementation  by Fuchs and Jacobs\cite{Fuchs}. Theorem 2 gives
this `rawness' a concrete meaning.  The term `raw' has
connotations of simplicity and unembellishment.  These
descriptions fit this implementation of a generalised measurement,
reflecting as they do the absence of an attempt to restore or
increase the purity of the state following acquisition of the
measurement outcome, which is captured by the majorization
relation (3.22).

It is instructive to compare and contrast Theorem 2 with a related
theorem due to Uhlmann\cite{Uhlmann1,Uhlmann2,Uhlmann3}.  This
states that the eigenvalues of two density operators ${\rho}_{1}$
and ${\rho}_{2}$ obey the majorization relation ${\mbox{\boldmath
${\lambda}$}}({\rho}_{2}){\prec}{\mbox{\boldmath
${\lambda}$}}({\rho}_{1})$ if and only if there exists a
probability distribution $p_{k}$ and unitary operators $U_{k}$
such that
\begin{equation}
{\rho}_{2}=\sum_{k}p_{k}U_{k}{\rho}_{1}U^{\dagger}_{k}.
\end{equation}
For a further proof and discussion of this theorem, see
Wehrl\cite{Wehrl}. Nielsen and Chuang\cite{NC} also give a
particularly direct proof whose sufficiency part parallels that of
the proof we have given of Theorem 2 above. It is obvious that Eq.
(3.28) is a valid quantum operation, indeed one which satisfies
(3.21). In fact, the sufficiency part of Uhlmann's theorem is
easily seen to follow from the sufficiency part of Theorem 2 in
the special case where $A_{k}=\sqrt{p_{k}}U_{k}$.

The necessity parts of Uhlmann's theorem and Theorem 2 are, on the
other hand, disjoint. In Uhlmann's theorem, the emphasis is on the
density operators.  It says that if ${\mbox{\boldmath
${\lambda}$}}({\rho}_{2}){\prec}{\mbox{\boldmath
${\lambda}$}}({\rho}_{1})$ then there must be a probability
distribution $p_{k}$ and unitary operators $U_{k}$ {\em which
depend on the initial and final density operators} and satisfy
(3.28). In contrast, the emphasis in Theorem 2 is on the quantum
operation, {\em which is independent of the density operators} and
makes a statement about the properties that a particular operation
must have if it is never to decrease mixedness for any density
operator.
\subsection{Majorization and deterministic transformations}
Only operations which satisfy condition (3.23) do not decrease
mixedness, in the sense quantified by majorization, for any state.
A well-known property of majorization is that if ${\mbox{\boldmath
${\lambda}$}}({\rho}_{2}){\prec}{\mbox{\boldmath
${\lambda}$}}({\rho}_{1})$, then
$S({\rho}_{2}){\geq}S({\rho}_{1})$, where $S({\rho})=-{\mathrm
T}{\mathrm r}({\rho}{\log}{\rho})$ is the von Neumann entropy. It
follows that if ${\rho}_{2}$ is at least as mixed as ${\rho}_{1}$
in the sense of majorization, then its von Neumann entropy is also
at least as high as that of ${\rho}_{1}$.

The von Neumann entropy has long been used to quantify mixedness,
in the sense of {\em disorder}, in quantum mechanics.  However,
with the advent of the noiseless coding theorems for classical and
quantum information transmission, it has acquired a further
significance as a measure of {\em information} which is directly
analogous to that of the Shannon entropy in classical information
theory. In this context, the density operator represents an
ensemble of pure states.  Consider two ensembles ${\cal
E}_{1}=\{q_{j},|{\psi}_{j}^{1}{\rangle}\}$ and ${\cal
E}_{2}=\{q_{j},|{\psi}_{j}^{2}{\rangle}\}$, where $q_{j}$ is the
{\em a priori} probability of both $|{\psi}_{j}^{1}{\rangle}$ and
$|{\psi}_{j}^{2}{\rangle}$. These ensembles have the density
operators
\begin{eqnarray}
{\rho}_{1}({\mathbf
q})&=&\sum_{j}q_{j}|{\psi}_{j}^{1}{\rangle}{\langle}{\psi}_{j}^{1}|,\\
{\rho}_{2}({\mathbf
q})&=&\sum_{j}q_{j}|{\psi}_{j}^{2}{\rangle}{\langle}{\psi}_{j}^{2}|.
\end{eqnarray}

The noiseless coding theorem for classical\cite{Ccoding}
(quantum\cite{Schumacher}) information with pure quantum states
implies that the maximum rate of asymptotically error-free
classical (quantum) information transmission using ensemble ${\cal
E}_{i}$ is $S({\rho}_{i}({\mathbf q}))$ bits (qubits) per signal.
However, suppose that we can transform ${\cal E}_{1}$ into ${\cal
E}_{2}$ with unit probability. If $S({\rho}_{2}({\mathbf
q}))>S({\rho}_{1}({\mathbf q}))$, then clearly these coding
theorems will be violated.  Such ensemble transformations, which
increase the von Neumann entropy, must be impossible, and lead us
to suspect that ensemble transformations giving rise to the
majorization relation ${\mbox{\boldmath
${\lambda}$}}({\rho}_{2}({\mathbf q})){\prec}{\mbox{\boldmath
${\lambda}$}}({\rho}_{1}({\mathbf q}))$ will also be impossible
(except in the trivial case where all of the equalities in (3.18)
are satisfied.)

The transformations we have in mind here are clearly the
deterministic transformations described in the preceding section.
The question is then: do the eigenvalues of ensemble density
operators whose constituent pure states are related by a
deterministic transformation obey {\em any} majorization relation?
The answer, as we will now see, is yes: it is precisely the
reverse of that considered in Theorem 2, which is highly
satisfactory in view of the above considerations.

\begin{theorem} Let $\{|{\psi}_{j}^{1}{\rangle}\}$ and
$\{|{\psi}_{j}^{2}{\rangle}\}$ be sets of $N$ pure states.
Consider the mixtures ${\rho}_{1}({\mathbf q})$ and
${\rho}_{2}({\mathbf q})$ defined by Eqs. (3.29) and (3.30).  If
there is a deterministic transformation ${\cal
D}:|{\psi}_{j}^{1}{\rangle}{\rightarrow}|{\psi}_{j}^{2}{\rangle}\;{\forall}\;j
$, then
\begin{equation}
{\mbox{\boldmath ${\lambda}$}}({\rho}_{1}({\mathbf
q})){\prec}{\mbox{\boldmath ${\lambda}$}}({\rho}_{2}({\mathbf
q})),
\end{equation}
for every a priori probability vector ${\mathbf q}$.
\end{theorem}
Prior to proving this, we note that to speak of majorization
relations, ${\rho}_{1}({\mathbf q})$ and ${\rho}_{2}({\mathbf q})$
must have the same number of eigenvalues.  This condition is
easily satisfied by `padding out' the spectrum with the lower
number of non-zero eigenvalues with zeroes so that the spectra of
both density operators are of equal size.\\

 \noindent{\bf Proof:} We start with the following
observation made by Jozsa and Schlienz\cite{Jozsa}. For the {\em a
priori} probability vector ${\mathbf q}$, we define the matrix
${\mathbf Q}=\{\sqrt{q_{j}q_{j'}}\}$. Then ${\rho}_{1}({\mathbf
q})$ has the same nonzero eigenvalues, with the same
multiplicities, as ${\mathbf Q}\circ{\mbox{\boldmath
${\Gamma}$}}_{1}$, and likewise with ${\rho}_{2}({\mathbf q})$ and
${\mathbf Q}\circ{\mbox{\boldmath ${\Gamma}$}}_{2}$.  To see why,
consider the entangled state of two systems, $a$ and $b$,
\begin{equation}
|{\Phi}{\rangle}=\sum_{j}\sqrt{q}_{j}|{\psi}_{j}{\rangle}_{a}{\otimes}|x_{j}{\rangle}_{b},
\end{equation}
where $\{|{\psi}_{j}{\rangle}\}$ may be either the set
$\{|{\psi}^{1}_{j}{\rangle}\}$ or $\{|{\psi}^{2}_{j}{\rangle}\}$,
and $\{|x_{j}{\rangle}\}$ is an orthonormal set. The purity of
this state implies that the eigenvalues of the reduced density
operators are the same for each subsystem.  We find that
\begin{eqnarray}
{\rho}_{a}&=&\sum_{j}q_{j}|{\psi}_{j}{\rangle}{\langle}{\psi}_{j}|,\\
{\rho}_{b}&=&\sum_{jj'}\sqrt{q_{j}q_{j'}}{\langle}{\psi}_{j}|{\psi}_{j'}{\rangle}|x_{j'}{\rangle}{\langle}x_{j}|,
\nonumber \\ &=&\sum_{jj'}\{({\mathbf Q}\circ{\mbox{\boldmath
${\Gamma}$}})^{T}\}_{j'j}|x_{j'}{\rangle}{\langle}x_{j}|,
\end{eqnarray}
where ${\mbox{\boldmath ${\Gamma}$}}$ is the Gram matrix of the
set $\{|{\psi}_{j}{\rangle}\}$. Eq. (3.34) tells us that the
elements of ${\rho}_{b}$ in the $\{|x_{j}{\rangle}\}$ basis give
the matrix $({\mathbf Q}\circ{\mbox{\boldmath ${\Gamma}$}})^{T}$,
where the superscript $T$ denotes the transpose.  Any Hermitian
matrix has the same nonzero eigenvalues as its transpose (with
corresponding eigenvectors being related by complex conjugation in
the standard basis.) So, we see that ${\rho}_{a}$ and ${\mathbf
Q}\circ{\mbox{\boldmath ${\Gamma}$}}$ have the same nonzero
eigenvalues.  This implies that
\begin{equation}
{\mbox{\boldmath ${\lambda}$}}({\rho}_{1}({\mathbf
q})){\prec}{\mbox{\boldmath ${\lambda}$}}({\rho}_{2}({\mathbf
q})){\Leftrightarrow}{\mbox{\boldmath ${\lambda}$}}({\mathbf
Q}\circ{\mbox{\boldmath ${\Gamma}$}}_{1}){\prec}{\mbox{\boldmath
${\lambda}$}}({\mathbf Q}\circ{\mbox{\boldmath
${\Gamma}$}}_{2})\;\;{\forall}\;\;{\mathbf q}.
\end{equation}

Consequently, we will be able to establish the majorization
relation (3.31) if we can establish that on the right hand side of
(3.35).  It turns out that the latter relation can be proven
rather straightforwardly using the following result
 obtained by Bapat and Sunder\cite{Bapat}: let
${\mathbf A}$ and ${\mathbf B}$ be $N{\times}N$ Hermitian
matrices.  If ${\mathbf A}{\geq}0$, and the diagonal elements of
${\mathbf A}$ are all equal to 1, then\cite{Footnote1}
\begin{equation}
{\mbox{\boldmath ${\lambda}$}}({\mathbf A}{\circ}{\mathbf
B}){\prec}{\mbox{\boldmath ${\lambda}$}}({\mathbf B}).
\end{equation}
Let us apply this relation, making the identifications:
\begin{eqnarray}
{\mathbf A}&=&{\mbox{\boldmath ${\Pi}$}},
\\ {\mathbf B}&=&{\mathbf
 Q}{\circ}{\mbox{\boldmath ${\Gamma}$}}_{2},
\end{eqnarray}
where ${\mbox{\boldmath ${\Pi}$}}$ is a positive, $N{\times}N$
matrix with diagonal elements equal to 1 and ${\mathbf B}$ is
easily shown to be Hermitian, indeed positive as a consequence of
the positivity of Gram matrices and projectors (${\mathbf Q}$
clearly being a projector) and Schur's product theorem\cite{Horn},
which states that the Hadamard product of two positive matrices is
also positive. Substituting these definitions into (3.36) gives
\begin{equation}
{\mbox{\boldmath ${\lambda}$}}({\mbox{\boldmath
${\Pi}$}}{\circ}({\mathbf Q}\circ{\mbox{\boldmath
${\Gamma}$}}_{2})){\prec}{\mbox{\boldmath ${\lambda}$}}({\mathbf
Q}\circ{\mbox{\boldmath ${\Gamma}$}}_{2}).
\end{equation}
We know from Eq. (2.11) that for a deterministic transformation,
there exists a positive matrix ${\mbox{\boldmath ${\Pi}$}}$ such
that ${\mbox{\boldmath ${\Gamma}$}}_{1}-{\mbox{\boldmath
${\Pi}$}}{\circ}{\mbox{\boldmath ${\Gamma}$}}_{2}=0$.  We can see
from this equation, or from the determinism condition together
with condition (1.b) of Theorem 1, that the diagonal elements of
${\mbox{\boldmath ${\Pi}$}}$ must be equal to 1. Making use of the
commutativity of the Hadamard product, we can easily see that, for
a deterministic transformation,

\begin{equation}
{\mathbf Q}{\circ}{\mbox{\boldmath
${\Gamma}$}}_{1}={\mbox{\boldmath ${\Pi}$}}{\circ}({\mathbf
Q}{\circ}{\mbox{\boldmath ${\Gamma}$}}_{2}).
\end{equation}
We can then substitute ${\mathbf Q}{\circ}{\mbox{\boldmath
${\Gamma}$}}_{1}$ into the left hand side of (3.39), giving the
majorization relation on the right hand side of (3.35). This
completes the proof.${\Box}$\\

A question of obvious importance whether or not the converse of
Theorem 3 is true, that is, whether or not satisfaction of the
majorization relation (3.31) is a sufficient condition for the
existence of a deterministic transformation ${\cal
D}:|{\psi}_{j}^{1}{\rangle}{\rightarrow}|{\psi}_{j}^{2}{\rangle}\;{\forall}\;j
$. At the time of writing, this question is open.  If it is ever
to be answered in the affirmative, then this could suggest an
interesting parallel between the theory of deterministic
transformations of sets of pure states, and that of deterministic
LOCC on pure, bipartite entangled states, which is covered by a
theorem due to Nielsen which we mentioned earlier. To be specific,
let $|{\Psi}_{1}{\rangle}$ and $|{\Psi}_{2}{\rangle}$ be a pair of
pure, bipartite entangled states, and ${\rho}_{1},{\rho}_{2}$ be
the corresponding reduced density operators for one of the
subsystems.  Then Nielsen's theorem\cite{Nielsen1} states the
necessary and sufficient condition for the existence of a
deterministic LOCC procedure which transforms
$|{\Psi}_{1}{\rangle}$ into $|{\Psi}_{2}{\rangle}$ is
\begin{equation}
{\mbox{\boldmath ${\lambda}$}}({\rho}_{1}){\prec}{\mbox{\boldmath
${\lambda}$}}({\rho}_{2}),
\end{equation}
The similarity between (3.31) and (3.41) is striking, especially
when we consider the fact that, in both contexts, the mixing,
whose non-increase is expressed by the appropriate majorization
relation, is related to a useful quantity or resource, rather than
simple disorder.  In the context of deterministic transformations
of sets of pure states, the degree of mixing can be intuitively
understood as expressing the distinguishability of the set of
states.  We feel that a further open problem, whose solution may
require that of the preceding one, is how one can make this
intuition quantitatively precise.

In the second context, that of deterministic LOCC entanglement
transformations, the degree of mixing relates to how entangled the
state is.  The fact that useful quantities such as entanglement
and distinguishability cannot increase under the appropriate kinds
of deterministic transformation, and that this fact can be
expressed by simple, similar majorization relations suggests that
both scenarios are related, and that this relationship could be
understood in terms of some broader, as yet unproposed unifying
framework.

\section{Discussion}
\renewcommand{\theequation}{4.\arabic{equation}}
\setcounter{equation}{0} In this paper, we have obtained some
general results relating to transformations of quantum states, and
associated probabilities or density operator eigenvalues, which
are closely related to and in some contexts can be interpreted as
probabilities. The main emphasis has been on transformations of
pure states. Probabilities play a essential role in quantum
mechanics in quantifying the likelihood of a particular
measurement outcome, given certain information about how the
system was prepared, namely its initial state vector or, more
generally, density operator.  This has been known since the early
days of quantum theory.  However, in recent decades, it has become
apparent, though a careful analysis of the postulates of quantum
mechanics and exploiting the possibilities afforded by
interactions between quantum systems, that the quantum formalism
permits more general measurements than those whose outcome
probabilities are obtained by direct application of Born's rule,
and where the resulting post-measurement states are obtained by
direct application of the von Neumann-Luders projection postulate.
Such measurements are known as generalised measurements.  The
formalism of quantum operations, which describes both aspects of
this general measurement process, has been of enormous interest
recently, especially due to its relevance to the developing field
of quantum information theory.

Since the early days of quantum theory,  it was recognised that
the measurement process is inextricably bound up with a
disturbance of the state of the system.  With the development of
generalised measurements, it has become recognised  that the large
disturbance associated with a sharp, von Neumann measurement is an
extremal case of a general trade-off between information and
disturbance\cite{FP,Fuchs}.  In this context, information is
treated as a `good' thing, while the associated disturbance is
considered to be an undesirable but unavoidable by-product.
However, in situations where we aim to tailor the disturbance to
produce a particular state, and where we wish to minimise the
probability of other transformations being carried out, it is
almost as though the conventional `morality' of the
information/disturbance trade-off is inverted.

Specific probabilistic transformations, such as cloning and
unambiguous state discrimination (which is a probabilistic
transformation of a non-orthogonal set into an orthogonal set)
have been studied in detail.   A further kind of transformation
which has been examined is a deterministic transformation, which
converts one set of pure states into another with unit
probability. However, probabilistic transformations, of which
deterministic transformations represent a limiting case, have not
previously been investigated in full generality.   To do so was
the objective of section II.  For pure state transformations, we
derived necessary and, when the initial states are linearly
independent, sufficient conditions for such a transformation to be
possible with given conditional probabilities for each of the
states.

Extending our analysis to cover more general quantum operations,
it is easily shown that the purity of states is not preserved in
general.  For the sake of simplicity, the probabilistic assumption
was removed and our emphasis shifted from selective to
nonselective operations.  This scenario is of considerable
practical importance since it applies to a quantum system whose
state we wish to control, deterministically, such as that of a
quantum computer, but which is subject to uncontrollable
influences such as that of the environment.

One of the most basic questions we can ask about such quantum
operations is: under what circumstances is the final state always
at least as mixed as the initial state, for every possible initial
state?  Quantifying the extent to which a state is mixed, at least
when the Hilbert space dimension exceeds 2, is non-trivial.
However, under certain circumstances, we can unambiguously compare
the degree of mixing of two quantum states for arbitrary quantum
systems; specifically, when the eigenvalues of one density
operator majorize those of the other.  The nontriviality of mixing
comparison is quantitatively captured by the fact that
majorization enforces only a partial order on equivalence classes
of density operators (with respect to unitary symmetry) which
allows for incomparable states.  We showed that a simple, elegant,
sufficient condition obtained by Bapat and Sunder is also
necessary.  We then
 showed that the eigenvalues of the source
density operators for initial and final pure state ensembles
related by a deterministic transformation obey the opposite
majorization relation.  In this context, mixing, rather than
characterising disorder, is related to the information content or
distinguishability of the ensemble, and this majorization relation
expresses the fact that such aspects of an ensemble cannot be
amplified and is perhaps in the same spirit as the no-cloning
theorem.  Indeed, it is quite simple to show that the strong form
of the no-cloning theorem, which states that it is impossible to
deterministically copy a set of non-orthogonal states, follows
from this majorization relation.

We noted the resemblance between this majorization relation and
that obtained by Nielsen in the context of LOCC entanglement
transformations.  The latter is a necessary and sufficient
condition for deterministic transformation of one pure, bipartite
entangled state into another.  The former is only known to be a
necessary condition for a deterministic transformation of one set
of pure states into another pure set.  We argued that if it can
also be shown to be sufficient, then there is the possibility that
deterministic LOCC and pure set transformations could be
incorporated within and understood in terms of a broader
encompassing framework.  This could lead to interesting insights
into the relationship between entanglement and distinguishability.

With this possibility in mind, let us consider the fact that the
majorization relation (3.31) implies that the quantities
\begin{equation}
{\mu}({\mathbf
q})=\sum_{r=1}^{k}{\lambda}^{\downarrow}_{r}\left[\sum_{j}q_{j}|{\psi}_{j}^{1}{\rangle}{\langle}{\psi}_{j}^{1}|\right],
\;\;\;\;\;\;1{\leq}k{\leq}N
\end{equation}
are non-increasing under any deterministic transformation ${\cal
D}:|{\psi}_{j}^{1}{\rangle}{\rightarrow}|{\psi}_{j}^{2}{\rangle}\;{\forall}\;j
$ for any set of final states $\{|{\psi}_{j}^{2}{\rangle}\}$.  Can
we refer to such quantities as `distinguishability monotones', by
analogy with the concept of entanglement monotones introduced by
Vidal\cite{Vidal}?  If so, then how are they related to operations
which distinguish between quantum states?  How are they related to
more general sets of distinguishability monotones?  Indeed, what
criteria are the necessary and sufficient conditions to qualify a
functional as being a distinguishability monotone, or measure? To
answer these questions, we would require a greater understanding
of the distinguishability of sets of pure quantum states,
comparable to that which we have of pure, bipartite entanglement.

\section*{Acknowledgements}
This work was supported by the UK Engineering and Physical
Sciences Research Council.


\begin{thebibliography}{99}
\bibitem{Footnote1}
The majorization relation (3.36) is actually a special case of the
following more general result obtained by Visick\cite{Visick}: let
${\mathbf A}$ and ${\mathbf B}$ be $N{\times}N$ Hermitian
matrices.  If ${\mathbf A}{\geq}0$ then ${\mbox{\boldmath
${\lambda}$}}({\mathbf A}{\circ}{\mathbf
B}){\prec}{\mbox{\boldmath ${\lambda}$}}(({\mathbf
1}{\circ}{\mathbf A}){\mathbf B})$.  This reduces to the
Bapat-Sunder majorization relation (3.36) in the special case
where all of the diagonal elements of ${\mathbf A}$ are equal to
1, in which case ${\mathbf 1}{\circ}{\mathbf A}={\mathbf 1}$.
\end{thebibliography}
\end{document}